\documentstyle[12pt]{article}
\begin{document}
\vskip 0.1in
\centerline{\Large\bf $1/R$ Curvature Corrections as the Source} 
\centerline{\Large\bf of the Cosmological Acceleration}
\vskip .7in
\centerline{Dan N. Vollick}
\centerline{Department of Physics and Astronomy}
\centerline{and}
\centerline{Department of Mathematics and Statistics}
\centerline{Okanagan University College}
\centerline{3333 College Way}
\centerline{Kelowna, B.C.}
\centerline{V1V 1V7}
\vskip .9in
\centerline{\bf\large Abstract}
\vskip 0.5in
Corrections to Einstein's equations that become important at small curvatures
are considered. The field equations are derived using a Palatini variation
in which the
connection and metric are varied independently. In contrast to the 
Einstein-Hilbert variation, which yields fourth order equations, the Palatini 
approach produces second order equations in the metric. The Lagrangian
$L(R)=R-\alpha^2/R$ is examined and it is shown that it leads to equations
whose solutions approach a
de Sitter universe at late times. Thus, the inclusion of $1/R$
curvature terms in the gravitational action offers an alternative explanation
for the cosmological acceleration.
\newpage
\section*{Introduction}
One of the most interesting aspects of modern cosmology concerns the 
acceleration of the cosmological expansion. Recent supernovae 
\cite{To1,Pe1,Ri1,Pe2} and
CMBR \cite{Be1,Ne1,Ha1,Ja1,La1,Me1} observations 
indicate that the expansion of the universe is
accelerating, contrary to previous expectations.
  
Most attempts to explain this acceleration involve the introduction of dark
energy, as a source of the Einstein field equations. The nature of
the dark energy is unknown but it behaves like a fluid with a large negative
pressure. One possible candidate for the dark energy is a very small
cosmological constant.
Another, possibly related, problem involves the existence of dark
matter. Observations of spiral galaxies, elliptical galaxies and galactic
clusters indicates that these objects contain a large amount of dark matter.
The difference between dark energy and dark matter is that dark matter clusters
with the visible matter and dark energy is more or less uniformly spread
throughout the universe.
  
Of course, one possibility is that we do not understand gravity on these 
large scales. Since dark energy and dark matter are needed to explain 
phenomena in regions of low curvature we can attempt to modify Einstein's
theory by adding corrections that become important when the curvature is
small (see \cite{Def1,Dva1,Dva2,Dva3} for other approaches that involve
modifications of Einstein's theory).
Recently two attempts \cite{Ca1,Cap1} to explain the cosmic acceleration
along these lines have been made. They involve adding a term proportional to
$1/R$ to the Einstein-Hilbert action and varying the action with respect to
the metric (see \cite{Kle1,Bra1} for other papers that consider non-polynomial
terms in the action). 
This approach leads to complicated fourth order equations that
can be simplified by performing a canonical transformation and introducing
a fictitious scalar field. It was shown in these papers that the modified 
field equations can produce the observed cosmological acceleration without the
need of dark energy.
  
In this paper I also consider a correction to the action that is proportional
to $1/R$, but I use the Palatini variational principle to derive the field
equations. In this approach $g_{\mu\nu}$ and $\Gamma^{\alpha}_{\mu\nu}$
are taken as independent variables. In Einstein gravity the variation with
respect to $\Gamma^{\alpha}_{\mu\nu}$ gives the usual relationship between
$\Gamma^{\alpha}_{\mu\nu}$ and the metric, i.e. $\Gamma^{\alpha}_{\mu\nu}$
is the Christoffel symbol associated with the metric $g_{\mu\nu}$. The
variation with respect to $g_{\mu\nu}$ gives $R_{\mu\nu}(\Gamma)-\frac{1}{2}
g_{\mu\nu}R(\Gamma)=-\kappa T_{\mu\nu}$, where $\kappa=8\pi G$. Thus, the
Palatini variation is equivalent to the Einstein-Hilbert variation in 
Einstein's theory. This is not the case however for other Lagrangians. In fact,
it has been shown \cite{Fe1,Fe2} that the Palatini variation gives the usual 
vacuum Einstein equations for generic Lagrangians of the form $L(R)$. This is 
to be contrasted to the purely metric variation that produces fourth order 
equations. If matter is included the Palatini variation still produces second
order equations, but the are no longer identical to Einstein's equations.
In this paper I show that the Palatini variation of the Lagrangian $L(R)=
R-\alpha^2/R$, where $\alpha$ is a constant, leads to field equations that 
give an accelerating universe at late times.

\section*{The Field Equations}
The field equations follow from the variation of the action
\begin{equation}
S=\int\left[ -\frac{1}{2\kappa}L(R)+L_M\right]\sqrt{g}d^4x
\end{equation}
where
\begin{equation}
R^{\alpha}_{\;\;\mu\nu\beta}=\partial_{\beta}\Gamma^{\alpha}_{\mu\nu}-
\partial_{\nu}\Gamma^{\alpha}_{\mu\beta}+\Gamma^{\lambda}_{\mu\nu}
\Gamma^{\alpha}_{\beta\lambda}-\Gamma^{\lambda}_{\mu\beta}\Gamma^{\alpha}_
{\nu\lambda}\; ,
\end{equation}
  
\noindent
$R_{\mu\nu}=R^{\alpha}_{\;\;\mu\alpha\nu}$, $R=g^{\mu\nu}R_{\mu\nu}$, 
$\kappa=8\pi G$ and
$L_M$ is the matter Lagrangian. Here we consider a Palatini variation of the
action, which treats $g_{\mu\nu}$ and $\Gamma^{\alpha}_{\mu\nu}$ as independent
variables. 
  
Varying the action with respect to $g_{\mu\nu}$ gives
\begin{equation}
L^{'}(R)R_{\mu\nu}-\frac{1}{2}L(R)g_{\mu\nu}=-\kappa T_{\mu\nu}
\label{fe}
\end{equation}
where $T_{\mu\nu}$ is the energy-momentum tensor and is given by
\begin{equation}
T_{\mu\nu}=-\frac{2}{\sqrt{g}}\frac{\delta S_M}{\delta g^{\mu\nu}}\; .
\end{equation}
Varying the action with respect to $\Gamma^{\alpha}_{\mu\nu}$ gives
\begin{equation}
\nabla_{\alpha}[L^{'}\sqrt{g}g^{\mu\nu}]-\frac{1}{2}\nabla_{\sigma}
[L^{'}\sqrt{g}g^{\sigma\mu}]\delta^{\nu}_{\alpha}-\frac{1}{2}\nabla_
{\sigma}[L^{'}\sqrt{g}g^{\sigma\nu}]\delta^{\mu}_{\alpha}=0 .
\end{equation}
By contracting over $\alpha$ and $\mu$ it is easy to see that this is
equivalent to
\begin{equation}
\nabla_{\alpha}[L^{'}(R)\sqrt{g}g^{\mu\nu}]=0.
\label{Chris}
\end{equation}
This equation can be solved for the connection using a similar approach to 
that used in general relativity. Alternatively, one can define a metric
$h_{\mu\nu}=L^{'}g_{\mu\nu}$ and it is easy to see that equation 
(\ref{Chris}) implies that the connection is the Christoffel symbol with
respect to the metric $h_{\mu\nu}$. A conformal transformation back to the
metric $g_{\mu\nu}$ gives (see \cite{Haw1} for the details on conformal
transformations)
\begin{equation}
\Gamma^{\alpha}_{\mu\nu}=\left\{
\begin{array}{ll}
\alpha\\
\mu\nu\\
\end{array}
\right\}+\frac{1}{2L^{'}}[2\delta_{(\mu}^{\alpha}\partial_{\nu)}L^{'}-g_{\mu
\nu}g^{\alpha\beta}\partial_{\beta}L^{'}]
\label{Chris2}
\end{equation}
where the first term is the Christoffel symbol with
respect to the metric $g_{\mu\nu}$. At first sight it might appear that this 
does not really define the connection since $L^{'}$ contains derivatives of the
connection. However, if we contract the field equation (\ref{fe}) we get
\begin{equation}
RL^{'}(R)-2L(R)=-\kappa T.
\label{alg}
\end{equation}
If this equation can be solved for $R=R(T)$, as we will assume here, then the
terms in (\ref{Chris2}) involving $L^{'}(R)$ 
can be expressed as derivatives of $T$. 
Since $T$ contains only the metric and not its derivatives the connection will
involve only first derivatives of the metric and the field equations will then
be second order in the metric $g_{\mu\nu}$. 
  
The Ricci tensor and Ricci scalar are given by 
\begin{equation}
R_{\mu\nu}=R_{\mu\nu}(g)-\frac{3}{2}(L^{'})^{-2}\nabla_{\mu}L^{'}\nabla_{\nu}
L^{'}+(L^{'})^{-1}\nabla_{\mu}\nabla_{\nu}L^{'}+\frac{1}{2}(L^{'})^{-1}
g_{\mu\nu}\Box L^{'}
\end{equation}
and
\begin{equation}
R=R(g)+3(L^{'})^{-1}\Box L^{'}-\frac{3}{2}(L^{'})^{-2}\nabla_{\mu}
L^{'}\nabla^{\mu}L^{'}
\end{equation}
where $R_{\mu\nu}(g)$ is the usual expression for $R_{\mu\nu}$ in terms of
$g_{\mu\nu}$ and $R=g^{\mu\nu}R_{\mu\nu}$.
  
If $T=0$ then  the solutions to equation (\ref{alg}) will be constants and
$h_{\mu\nu}$ will be equal to $g_{\mu\nu}$ times a (positive) constant.
This implies that
$R_{\mu\nu}=R_{\mu\nu}(g)$, and $R=R(g)$. Thus, in a vacuum the field
equations will reduce to the Einstein field equations 
with a cosmological constant for a generic $L(R)$
(see \cite{Fe1,Fe2} for the vacuum case).
The field equations (\ref{fe}) can be written in the Einstein form 
\begin{equation}
R_{\mu\nu}(g)-\frac{1}{2}R(g)g_{\mu\nu}=-\kappa S_{\mu\nu}
\end{equation}
with a modified source $S_{\mu\nu}$ given by
\begin{equation}
S_{\mu\nu}=(L^{'})^{-1}T_{\mu\nu}-\frac{1}{\kappa}\left[\frac{3}{2}(L^{'})^{-2}
\nabla_{\mu}L^{'}\nabla_{\nu}L^{'}-(L^{'})^{-1}\nabla_{\mu}\nabla_{\nu}L^{'}
-\frac{1}{2}g_{\mu\nu}(L^{'})^{-1}\Box L^{'}+\frac{1}{2}\left( R-\frac{L}
{L^{'}}\right) g_{\mu\nu}\right].
\end{equation}
  
Now consider the Lagrangian
\begin{equation}
L(R)=R-\frac{\alpha^2}{3R}
\end{equation}
where $\alpha$ is a positive constant with the same dimensions as $R$ and 
the factor of 
3 is introduced to simplify future equations. The field equations for this
Lagrangian are
\begin{equation}
\left[ 1+\frac{\alpha^2}{3R^2}\right] R_{\mu\nu}-\frac{1}{2}\left[ R
-\frac{\alpha^2}{3R}\right] g_{\mu\nu}=-\kappa T_{\mu\nu}\; .
\label{fe2}
\end{equation}
Contracting the indices gives
\begin{equation}
R^2-\kappa TR-\alpha^2=0 
\end{equation}
and the solution to this algebraic equation is
\begin{equation}
R=\frac{1}{2}\left[\kappa T\pm\sqrt{\kappa^2T^2+4\alpha^2}\right].
\label{sol}
\end{equation}
For large $|T|$ we expect the above to reduce to $R=\kappa T$, which follows
from the Einstein field equations. Thus, if $T>0$ we need to select the 
positive sign and if $T<0$ we need to select the negative sign. In a universe
filled with an ideal fluid $T=-(\rho-3P)$, so that $T<0$ if the dominant
energy condition holds (i.e. $\rho>0$ and $\rho\geq 3|P|$). Thus we have
\begin{equation}
R=\frac{1}{2}[\kappa T-\sqrt{\kappa^2T^2+4\alpha^2}]\; .
\end{equation}
The vacuum solution is
\begin{equation}
R=-\alpha 
\label{R}
\end{equation}
so that at late times, as $T^{\mu\nu}\rightarrow 0$, the universe will 
approach a de Sitter spacetime and the expansion of the universe will 
accelerate. 
     
From equations (\ref{fe2}) and (\ref{sol}) we see that the field equations 
reduce to the Einstein equations if $|\kappa T|>>\alpha$. Thus, in a
dust filled universe the evolution will be governed by the Einstein field
equations at early times. Eventually the corrections to the equations of motion
will become important and the universe will make a transition to a de Sitter 
universe at late times. To match the observations of the cosmological
acceleration we must take $\alpha\sim 10^{-67}\; (eV)^2\sim 10^{-53}\; m^{-2}$.
Note that in a vacuum we get the Einstein field equations plus a small
cosmological constant, so that this theory will pass all the solar system 
tests that general relativity has passed. It is also interesting to note
that in a radiation dominated universe $T=0$ so that the dynamics is not
governed by the Einstein's equations even at large curvature. As we will see
below (see equation (\ref{fe3})) the equations of motion are the Einstein field
equations with a cosmological constant and a modified Newton's constant.
  
Now consider the evolution of a universe with metric
\begin{equation}
ds^2=-dt^2+a(t)^2[dx^2+dy^2+dz^2]
\end{equation}
at late times, when $\alpha>>\kappa T$. In this regime
\begin{equation}
R_{\mu\nu}\simeq R_{\mu\nu}(g)+\frac{1}{L^{'}}\nabla_{\mu}\nabla_{\nu}L^{'}
+\frac{1}{2L^{'}}g_{\mu\nu}\Box L^{'},
\end{equation}
\begin{equation}
R\simeq\frac{1}{2}\kappa T-\alpha,
\end{equation}
\begin{equation}
L\simeq -\frac{2}{3}\alpha \left[1-\frac{\kappa T}{\alpha}\right]
\end{equation}
\begin{equation}
L^{'}\simeq\frac{4}{3}\left[1+\frac{\kappa T}{4\alpha}\right],
\end{equation}
and the field equations are
\begin{equation}
R_{\mu\nu}\simeq-\frac{1}{4}\alpha g_{\mu\nu}-\kappa\left[\frac{3}{4}T_{\mu\nu}-
\frac{5}{16}Tg_{\mu\nu}\right]
\label{fe3}
\end{equation}
The matter in the present universe can be approximated by dust 
with $T=\rho_0/a^3$, where $\rho_0$ is a constant. 
The nonvanishing components of the Ricci tensor are
\begin{equation}
R_{tt}=\frac{3\ddot{a}}{a}+\frac{9\kappa\rho_0}{8\alpha a^3}\left[
\frac{\ddot{a}}{a}-3\left(\frac{\dot{a}}{a}\right)^2\right]
\end{equation}
and
\begin{equation}
R_{ij}=-\left[a\ddot{a}+2\dot{a}^2+\frac{3\kappa\rho_0}{8\alpha a}\left(
\frac{\ddot{a}}{a}+\frac{\dot{a}^2}{a^2}\right)\right]\delta_{ij} .
\end{equation}
At late times the universe will almost be in
a de Sitter phase and we can take
\begin{equation}
a(t)=e^{Ht}+b(t)
\end{equation}
where $|b(t)|<<e^{Ht}$ and $\alpha=12H^2$. 
To lowest order in $b$ the field equations are
\begin{equation}
\ddot{b}-H^2b=-\frac{\kappa\rho_0}{12}e^{-2Ht}
\end{equation}
and 
\begin{equation}
\ddot{b}+4H\dot{b}-5H^2b=\frac{\kappa\rho_0}{4}e^{-2Ht}.
\end{equation}
Subtracting these two equations gives the first order equation
\begin{equation}
\dot{b}-Hb=\frac{\kappa\rho_0}{12H}e^{-2Ht}\; ,
\end{equation}
and the particular solution to this equation is
\begin{equation}
b(t)=-\frac{\kappa\rho_0}{36H^2}e^{-2Ht}\; .
\end{equation}
Thus, at late times the universe approaches a de Sitter spacetime exponentially
fast. This behaviour is analogous to the cosmic no hair theorem for fourth
order gravity discussed by Kluske and Schmidt \cite{Kl1}.
\section*{Conclusion}
Using a Palatini variation the field equations for a  nonlinear gravitational
Lagrangian coupled to matter were found. The vacuum field equations are the
Einstein equations with a cosmological constant. Thus, at late times as
$T_{\mu\nu}\rightarrow 0$ our universe will approach a de Sitter spacetime.
The inclusion of matter gives field equations that differ Einstein's 
equations. Using these equations it was shown that the approach to de Sitter
space is exponentially fast when the $1/R$ term dominates.
Thus, the inclusion of non-polynomial curvature terms in the gravitational
action offers an alternative explanation for the cosmological acceleration.

\end{document}